# INTERSTELLAR TURBULENCE

D. Falceta-Gonçalves [1]

RESUMEN

O meio interestelar (ISM) é um sistema complexo, em várias fases, onde a história das estrelas ocorre. Os processos de formação e morte das estrelas estão fortemente acoplados à dinâmica do meio. Os movimentos caóticos e difusivos observados no meio interestelar caracterizam a turbulência local. A compreenso da turbulência é crucial para o entendimento dos processos de formação estelar e o feedback de massa e energia ao meio a partir das estrelas evoluidas. Campos magnéticos também permeiam o meio interestelar fazendo dessa tarefa ainda mais complicada. Neste trabalho, eu reviso brevemente as principais observações e a caracterização da turbulência no meio interestelar. Também, mostro os principais resultados teóricos e de simulações numéricas que podem ser usados para construir observáveis, e comparo essas previsões a algumas observações.

ABSTRACT

The Interstellar Medium (ISM) is a complex, multi-phase system, where the history of the stars occurs. The processes of birth and death of stars are strongly coupled to the dynamics of the ISM. The observed chaotic and diffusive motions of the gas characterize its turbulent nature. Understanding turbulence is crucial for understanding the star-formation process and the energy-mass feedback from evolved stars. Magnetic fields, threading the ISM, are also observed, making this effort even more difficult. In this work, I briefly review the main observations and the characterization of turbulence from these observable quantities. Following on, I provide a review of the physics of magnetized turbulence. Finally, I will show the main results from theoretical and numerical simulations, which can be used to reconstruct observable quantities, and compare these predictions to the observations.

## 1. INTRODUCTION

Turbulence is ubiquitous in the Universe, from scales of few centimeters in the water running in the sink of our homes, to kiloparsecs at the flowing plasmas in the intergalactic medium (Falceta-Gonçalves et al. 2010). The classical work of Kolmogorov in the 40's have provided a fundamental theoretical basis for the study of the dynamics of turbulence. In a very simplistic picture, the turbulence is characterized by chaotic motions in a fluid, resulting in diffusion of matter and decay of the kinetic energy from large to smaller scales, which end up killed by the fluid viscosity with the release of heat. But few questions remain: how does the energy cascades from the large to the smaller scales? or does this picture changes when the fluid is a plasma threaded by magnetic fields?

In order to answer the first question, Kolmogorov assumed that the turbulence in incompressible fluids should obey the principles of *i- homogeneity, ii- isotropy, iii- scale invariance* and *iv- locality*. By "locality" one should understand wave-wave interaction at a local wavenumber at Fourier space. In this sense, the large scale eddy ($k_1$) decays into smaller eddies $k_2$, being $k_2 = 2k_1$. The energy transfer rate at a given scale $l$ is given by $\dot{\epsilon} \sim \delta v_l^2/\tau_l$, being $\tau_l \sim l/\delta v_l$. Therefore, $\delta v_l \sim (\dot{\epsilon}l)^{1/3}$. Regarding the power spectrum, if the density is a passive scalar we can write $\delta v_l^2 \sim \int_k^\infty E(k')dk'$, with $k = 2\pi/l$, resulting in the well-known Kolmogorov's spectrum $E(k) \sim \dot{\epsilon}^{2/3} k^{-5/3}$.

In the presence of magnetic fields the turbulence cascades quite differently. The main issue is that the eddies are compressed by the field lines in the perpedicular direction only. In the model presented by Goldreich & Sridhar (1995) the timescales for alfvenic and sonic motions are assumed to be similar, i.e. $\tau_A \sim \tau_s$, which corresponds to $l_\parallel/l_\perp \sim c_A/\delta v_l$. Using the scaling relation of $\delta v_l$ one obtains $l_\parallel \propto l_\perp^{2/3}$. This result shows that the turbulence in magnetized plasmas is *anisotropic*, and that this anisotropy is larger at smaller scales.

Kolmogorov's scalings also fail for compressible turbulence. In the case of supersonic flows the shocks are responsible for energy exchanges that results also in compression of the fluid. In this sense, the specific energy of a fluid element depends not exclusively on its velocity but also on its density. If the energy transfer rate is written as $\dot{\epsilon} \sim \rho \delta v^2/(l/\delta v) \sim \rho \delta v^3/l$, one may introduce the variable $\delta u \equiv \rho^{1/3} \delta v$. The power spectrum of $\delta u$ develops a standard Kolmogorov's slope (i.e. $\alpha = -5/3$), but $\alpha \sim -2$ for $\delta v$.

---

[1] Escola de Artes, Ciências e Humanidades, Universidade de São Paulo - Rua Arlindo Bettio 1000, CEP: 03828-000, São Paulo, Brazil (dfalceta@usp.br).

## 2. THE TURBULENT ISM

In the past few decades many important observational studies have proven the turbulent nature of the interstellar medium (ISM), being i) the distribution of gas (or column density maps), and ii) the dispersion of line profiles, the two major indicators of it.

Armstrong, Rickett & Spangler (1995) presented the seminal work in which they used the scintilation of background radiation by the turbulent flows in the ISM in order to obtain the density spectrum along the line of sight. Assuming that the density may be considered a passive scalar - it is true for the warm and diffuse part of the ISM - the power spctrum may also be obtained. Their impressive result was a single Kolmogorov slope ranging for more than 6 orders of magnitudes in lengthscale. The data from the density distribution in this case corroborated with the previous findings of Larson (1981). Larson compiled the velocity dispersion distributions observed for several molecular clouds and determined the empirical scaling law $\delta v \propto l^a$, with $a \sim 0.3 - 0.5$ (on should remember that Kolmogorov's theory predicts $a = 1/3$, and for supersonic turbulence $a \sim 1/2$), as shown in Figure 1 (top).

However, in the past decade new and more sensitive observations of molecular clouds have shown a different picture. When dense cores are plotted together with the data compiled by Larson a less coherent scaling relation appears. The bottom image of Figure 1 (extracted from Ballesteros-Paredes et al. 2010) shows that the smaller scales present a broad range of observed dispersions of velocities, and that the Larson relation is not followed by all clouds. Ballesteros-Paredes et al. (2010) interpreted these observations as an indication that these cores were actually contracting. The gravitational collapse would result in an increase of the velocity dispersion at the scales of the denser cores.

Magnetic fields can also be used in order to study the ISM turbulence. Polarization maps - in the optical for dust absorption, and in the infrared by dust emission - are useful in unreaveling the topology of the magnetic field in the plane of sky. The perturbations in the general trend of the field lines, i.e. the uniform component, are usually associated to the turbulent motions in the plasma. In the Chandrasekhar-Fermi approximation the equipartition between the magnetic perturbation and the turbulent kinetic energy is assumed. In this case it is possible to estimate either the magnetic pressure or the dispersion of velocity once the other is known.

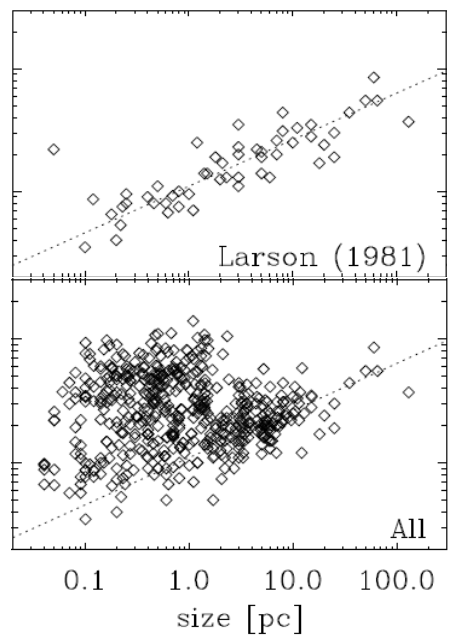

Fig. 1. Dispersion of velocity vs lengthscale for typical large molecular clouds (top) - original Larson relation - and for all structures, including the dense massive cores (bottom) (extracted from Ballesteros-Paredes 2010).

### 2.1. *Results from numerical simulations*

One of the main difficulties faced by observers in studying turbulence from the observations is its three dimensional nature. Observations are either 1-D, when projections along the line of sight are considered (e.g. dispersion of velocities, scintilation of background radiation), or 2-D when maps in the plane of the sky are used (e.g. column density statistics and polarization maps). The conversion from the observed quantities to the actual physics of turbulence is not straightforward (see Burkhart et al. 2009). In this sense numerical simulations represent a powerful tool. It is possible to simulate three dimensional turbulent flows and reconstruct observables from these cubes, considering any position for the observer.

In Falceta-Gonçalves, Lazarian & Houde (2010) we calculated synthetic observations of line profiles and presented dispersion of velocity *vs* lengthscale relationship for different turbulent regimes (Figure 2), being subsonic (left) and supersonic (right). It is clear in these calculations, even more in the supersonic case, that the range of velocity dispersions at small scales is very large. Is it gravity the main cause for that? Actually, these simulations were MHD, isothermal, non selfgravitating models. The main issue is that the "small scale" in the plot does not truly corresponds to the smaller turbulent cells. Since the line profiles are obtained integrating along the line of sight, it is obtained in a volume $V = l \times l \times L$, being $L$ the largest scale. Within the LOS, many small dense structures - each being subject to a drift velocity of the order of the largest scale amplitude - are intercepted and contribute to the total dispersion.

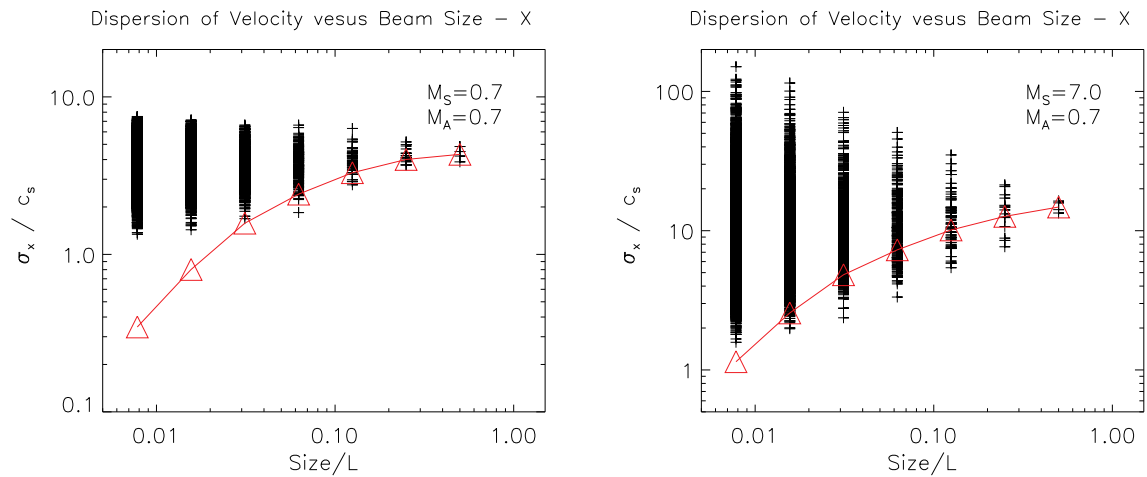

Fig. 2. Dispersion of velocity vs lengthscale, obtained for all LOS's with beam sizes $l \times l$, for two 3D-MHD numerical simulations. *Left*: subsonic/subalfvenic and *right*: supersonic/subalfvenic (extracted from Falceta-Gonçalves, Lazarian & Houde 2010). The line corresponds to the actual three dimensional scaling, i.e. $l^3$.

The origin of the condensations is related to the turbulence sonic Mach number. It reveals that shocks are the most important mechanism in generating dense structures in the ISM (Ballesteros-Paredes, Vázquez-Semadeni & Scalo 1999).

Regarding the magnetic fields, numerical simulations have shown that even subalfvenic turbulence, at large scales, may present superalfvenic flows at smaller scales (Falceta-Gonçalves, Lazarian & Kowal 2008). When shocks occur predominantly perpendicular to the mean field lines the density increases but the magnetic pressure not. Within the denser regions, the alfven speed (as it scales with $\rho^{-1/2}$) decreases and may become smaller than $\delta v_l$. The superalfvenic turbulence at the small scales is responsible for the bending of the field lines and, e.g. the decrease in the polarization degree, as observed in dense molecular cores.

## 3. IMPLICATIONS ON STAR FORMATION

It is well stablished that the ISM turbulence plays a key role in the star formation process, together with the magnetic field. In the same time, stellar energy feedback (mostly SNe) triggers the turbulence at large scales. Therefore, a full understanding of the turbulence cascade and the evolution of magnetic fields in turbulent flows is mandatory.

Uniform magnetic field lines are able to prevent the gravitational collapse of a cloud in the perpendicular direction, though not efficient in preventing the collapse as a whole. Alfvenic motions within the collapsing clouds may be effective in slowing down the inward fluxes of gas and, therefore, reduce the star formation efficiency (Falceta-Gonçalves, de Juli & Jatenco-Pereira 2003).

Another important issue is the survival of the dense cores to the turbulence itself. In one hand, the turbulent motions result in shocks and the formation of the dense clouds. In the other hand, the turbulence is responsible for the diffusion of the gas and the disruption of cloudlets and cores. How long can a star forming region survive?

From the fragmentation theory for star formation, a large molecular cloud gets gravitationally unstable and contracts. During the contraction its density gets larger and larger resulting in a decrease of the Jeans mass. Parts of this cloud with masses larger than the new critical mass "dettach" from the whole cloud and starts to collapse by its own. The process occurs until the optical depth of the cloud becomes much larger than unity. The last fragmented cores represent the final stage of a prestellar cloud.

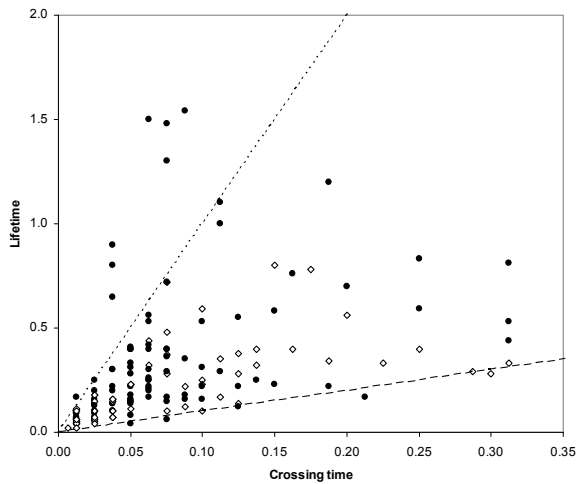

Fig. 3. Clump lifetimes given in terms of the sonic crossing time. Dashed line corresponds to $t_{\text{life}} = t_{\text{cross}}$ and dotted line corresponds to $t_{\text{life}} = 10 t_{\text{cross}}$. (extracted from Falceta-Gonçalves & Lazarian 2011, submitted)

The fragmentation and collapse of a cloud is, in general, believed to be fast. Much faster than other typical timescales of the system. Indeed, the ratio between the free-fall time and the turbulent turnover timescale for a Giant Molecular Cloud (GMC) is:

$$\frac{t_{\text{ff}}}{t_{turb}} \simeq \frac{c_{\text{s}}}{l(G\rho)^{1/2}} \sim 0.1. \qquad (1)$$

However, a typical 1pc sized dense core within a molecular cloud presents $t_{\text{ff}}/t_{turb} \sim 10$, i.e. the turbulence is able to fully develop during the core collapse. In order to test the survival of clumps to the turbulent diffusion we performed a standard, $512^3$ resolution, 3D MHD numerical simulation with $M_{\text{s}} \sim 7$, and identified and followed clumps for a long time. In Figure 3 we present the core lifetimes vs the sonic crossing time for all identified overdensity structures (clumps) in the run. It is clear that most of the detected clumps survive more than one crossing time, but only few can survive more than 10 crossing times (similar to the values found by Galván-Madrid et al. 2007). The turbulent diffusion is able to disrupt a clump before it fragments and collapses even further. Less than 5% of the clumps presented lifetimes larger than 10 crossing times. This result presents a possible explanation for the decreased star formation efficiency, even in regions where there is no massive star or any other source of strong radiation.

A rough estimate of the clump lifetimes is obtained assuming that all the gas shocking on a clump, with size $l$ and density $\rho_l$, is trapped:

$$\rho v_{\text{L}} t_{\text{life}} \sim \rho_l l; \qquad (2)$$

where $v_{\text{L}}$ is the large scale velocity and $\rho$ is the density of the unperturbed gas. This results in:

$$t_{\text{life}} \sim \frac{\rho_l l}{\rho v_{\text{L}}}. \qquad (3)$$

Equation 3 shows that the lifetime of a small scale clump depends on the large scale turbulent eddies as well. Also, provided that the column density may be obtained by $\Sigma \sim \rho_l l$, we get the scaling $t_{\text{life}} \propto \Sigma$.

D.F.G. thank the financial support of the Brazilian agency FAPESP (No. 2009/10102-0)

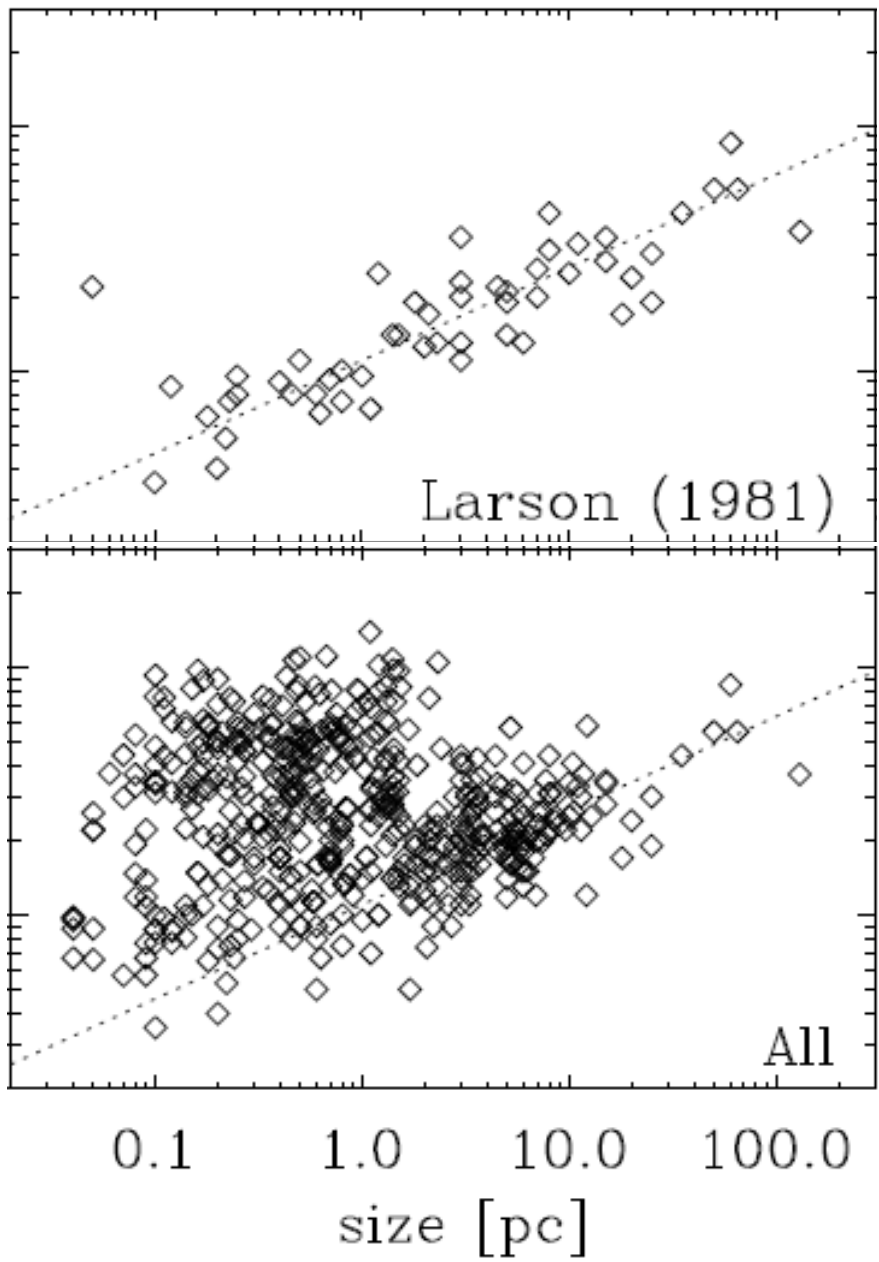

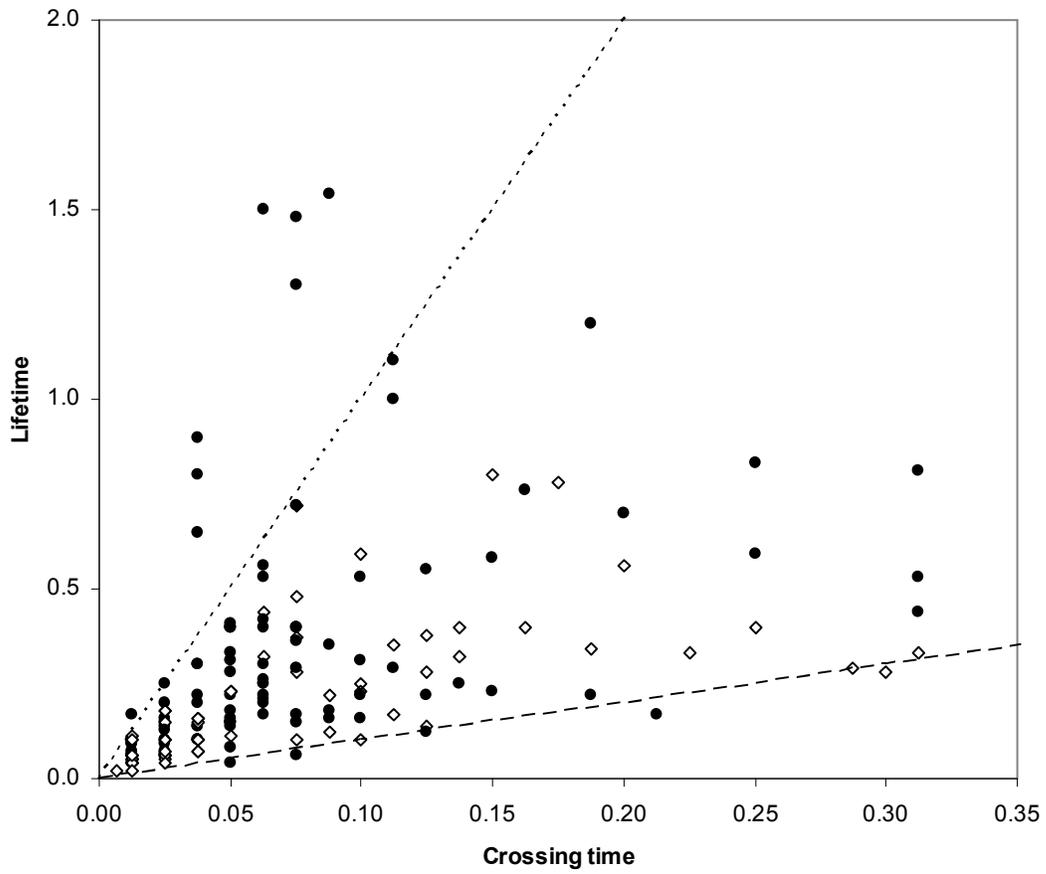

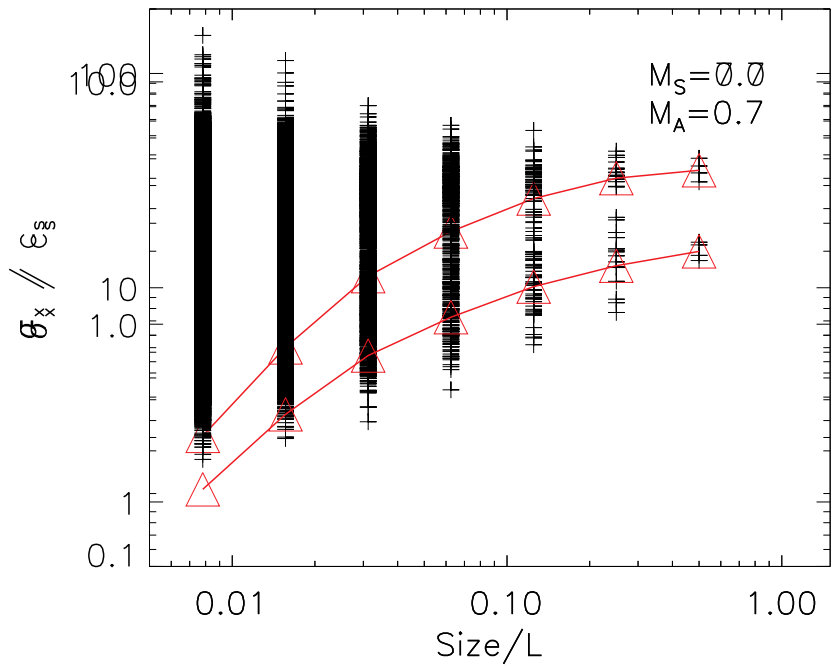